\documentclass[a4paper]{article}

\usepackage[english]{babel}
\usepackage[utf8x]{inputenc}
\usepackage[T1]{fontenc}
\usepackage{booktabs}
\usepackage{titlesec}
\usepackage{titling}

\titleformat{\paragraph}
{\normalfont\normalsize\bfseries}{\theparagraph}{1em}{}
\titlespacing*{\paragraph}
{0pt}{3.25ex plus 1ex minus .2ex}{1.5ex plus .2ex}

\usepackage[a4paper,top=3cm,bottom=2cm,left=3cm,right=3cm,marginparwidth=1.75cm]{geometry}

\usepackage{amsmath,amsfonts,array,subcaption,changepage}
\usepackage{graphicx,float}
\usepackage[colorinlistoftodos]{todonotes}
\usepackage[title]{appendix}
\usepackage[colorlinks=true, allcolors=blue]{hyperref}
\DeclareMathOperator*{\argminA}{arg\,min}

\title{\textbf{Self-Folding Metasheets:\\ The Optimal Pattern of Strain of Miura-Ori Folded State}}
\author{\textbf{Ling Lan}\\ll3178@nyu.edu\\ New York University\and \makebox[.9\textwidth]{Mentor: Miranda Holmes-Cerfon}\\holmes@cims.nyu.edu\\ New York University}
\date{}

\begin{document}
\maketitle

\begin{abstract}

Self-folding origami has emerged as a tool to make functional objects in material science. The common idea is to pattern a sheet with creases and activate them to have the object fold spontaneously into a desired configuration. This article shows that collinear quadrilateral metasheets are able to fold into the Miura-Ori configuration, if we only impose strain on part of their creases. In this study, we define and determine the \textbf{optimal pattern of strain (OPS)} on a collinear quadrilateral metasheet, that is the pattern of minimum ``functional'' creases with which the self-folding metasheet can fold into Miura-Ori state stably. By comparing the energy evolution along the folding pathway of each possible folded state under OPS, we conclude that the energy predominance of the desired Miura-Ori pathway during the initial period of time accounts for why the OPS works. Furthermore, we measure the projected force of the OPS on the intial flat metasheet and give insights on how to determine the OPS using only local information of the initial flat state.

\end{abstract}

\section{Introduction}\label{sec:intro}

One challenge in material science is to make functional objects at the nanoscale. These objects are far too small to build by hand, and self-folding origami has emerged as a tool for designing these three-dimensional structures from flat films \cite{Pandey,Miskin,Plucinsky}. Besides its great potential for the manufacture of complicated geometries and devices, self-folding origami opens up a number of research directions, such as origami design, the foldability and mechanics of origami, etc. Stern, Pinson and Murugan \cite{Menachem} discussed the complexity of refolding a previously folded sheet of paper and they provide fundamental limits on the programmability of energy landscapes in sheets. Chen and Santangelo \cite{Bryan} investigated the branches of a generic, triangulated origami crease pattern. Waitukaitis, Menaut, Chen, and Hecke \cite{Scott} defined the origami multistability and proved that rigid, degree-four vertices are generically multistable. The criteria of choosing available Mountain and Valley assignments in the Depth-First-Search algorithm introduced in subsection \ref{subsec:DFS} is based on their discussion on the foldability of a degree-four vertex. 

An origami structure is a system of rigid flat plates jointed pairwise by hinges (or creases). A mathematical interpretation of origami structures is provided more precisely in subsection \ref{subsec:defn}. The network formed by the creases and their junctions is called the crease pattern. Based on different assignments of Mountain and Valley options, an origami structure can take on a variety of configurations in 3D space. 

Santangelo \cite{Christian} discussed that there are basically two approaches to self-fold a crease. One way begins with a thin sheet of prestressed polymer glass. When heated locally, the release of the prestress causes the material to shrink locally and fold, if the origami structure is designed properly. A second approach starts with multilayer sheets of different materials. As the temperature changes, one surface expands to a greater extent than the other and the strip bends to the predetermined state. These methods all provide good control for producing folded structures. However, rather than modifying the mechanisms of folding, can we simplify the imposed pattern of strain that forces a flat sheet to buckle into a desired configuration?

In section \ref{sec:model}, we build up a stochastic model of a self-folding origami. The widely-used model of folding a self-folding origami imposes strain on all the creases in its crease pattern. However, when we simulate the folding process, we find that quadrilateral metasheets are able to fold into the Miura-Ori configurations, if we only impose strain on part of their creases. In section \ref{subsec:MO}, we introduce Miura-Ori, which is an origami structure that receives increasing attention for its negative Poisson's ratio and its high degree of symmetry folding in its periodicity \cite{Levi,Matthew}. The crease patterns of the Miura fold form a tessellation of the surface by parallelograms, that is a collinear quadrilateral metasheet. Here, we define the \textbf{optimal pattern of strain (OPS)} on a collinear quadrilateral metasheet to be the pattern of minimum ``functional'' creases with which the self-folding metasheet can fold into Miura-Ori state stably. 

It is natural to ask our first question: \textbf{does such an OPS exist?} In section \ref{sec:results1}, we introduce how we determine OPS numerically and we show that the OPS we find truely fold a flat metasheet into Miura-Ori state through stability tests. We also show that the stability of the OPS, i.e., the probablity of folding into Miura-Ori, not only depends on the choice of OPS, but also depends on the Miura-Ori angle and the noise during the folding process. Then we come up with our second question: \textbf{why the OPS can truely fold a flat metasheet into Miura-Ori state?} To answer this question, in section \ref{sec:results2}, we firstly figure out the available configurations under the OPS using a Depth First Search algorithm. Then we compare the energy evolution along each possible folding pathway and conclude that the energy predominance of the desired Miura-Ori pathway during the initial period of time accounts for why OPS works. Till now, we determine the OPS by looking at the probablity of success after a long period of time. But \textbf{can we determine the OPS using only local information of the initial flat state?} In section \ref{sec:results3}, we firstly build up the rigidity matrix of an initial flat metasheet and compute the projected force of the OPS on the flat metasheet. Although our method does not provide an OPS directly, it gives insights on how to determine OPS for Miura-ori configurations. Future directions of this research project are discussed in section \ref{sec:conclu}.

\section{Math Modeling and Miura-Ori Crease Pattern} \label{sec:model}

The goal of this section is to introduce a stochastic and computational model for self-folding origami structures, upon which the subsequent sections are built. Firstly, in section \ref{subsec:defn}, we introduce the definition and the energy function of an origami structure. Then, in section \ref{subsec:ODE}, we will explain how we model the folding process of an origami structure by finding the local minimum of its energy function. Finally, in section \ref{subsec:robust}, we test the robustness of our method and explain the modifications that make the numerical model more robust. To complete the background introduction, in section \ref{subsec:MO}, we introduce the Miura-Ori crease pattern, where all the further analysis in section \ref{sec:results} are based on.

\begin{subsection}{Definition of an Origami Structure and its Energy Function} \label{subsec:defn}

An origami structure contains the information of its vertices, edges, faces, and the angles between adjacent faces. Now, we will explain the model of each part and how the corresponding coefficients represent the properties of the material.

\begin{figure}[H]
 \centering
 \includegraphics[width=0.3\textwidth]{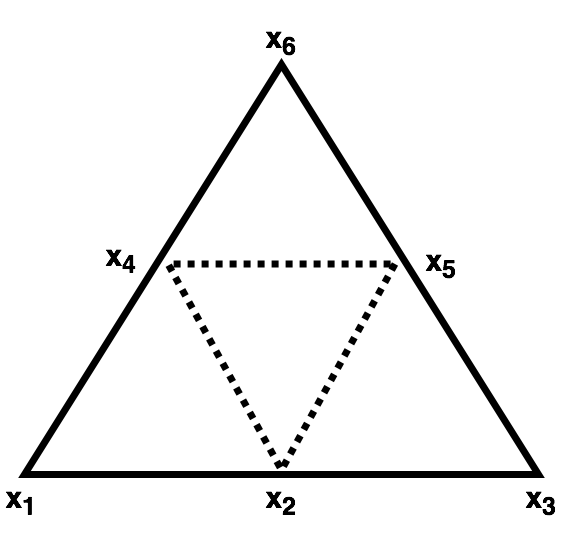}
 \caption{A flat origami structure of a tetrahedron.}
 \label{figure: tetrahedron}
\end{figure}

\begin{itemize}
\item $X:=\{\mathbf{x}_i\}_{i=1\cdots n}$ is the coordinates of the $n\in\mathbb{N}^+$ vertices. Notice that each configuration is uniquely represented by vertex coordinates. 
\item $G:=\{(\mathbf{x}_i,\mathbf{x}_j)\}$ is the set of edges with predetermined fixed length. Each pair of vertices $(\mathbf{x}_i,\mathbf{x}_j)$ in the set is connected and the order of vertices has no influence. For example, a flat origami structure that is able to fold into a tetrahedron shown in Figure \ref{figure: tetrahedron} has an edge set $\{(\mathbf{x}_1,\mathbf{x}_2),(\mathbf{x}_2,\mathbf{x}_3),(\mathbf{x}_1,\mathbf{x}_4),(\mathbf{x}_2,\mathbf{x}_4),(\mathbf{x}_2,\mathbf{x}_5),(\mathbf{x}_3,\mathbf{x}_5),(\mathbf{x}_4,\mathbf{x}_5),(\mathbf{x}_4,\mathbf{x}_6),(\mathbf{x}_5,\mathbf{x}_6)\}$.
Each edge $e_{ij}$ is modeled by a spring with a resting length $\overline{l_{ij}}$ and a stiffness coefficient $0<k_{ij}<1$. When $k_{ij}\to 1$, $l_{ij}$, the actual length of $e_{ij}$, is more reluctant to change; and when $k_{ij}\to 0$, $e_{ij}$ is easy to be enlarged or squeezed.
\item $A:=\{(\mathbf{x}_h,\mathbf{x}_i,\mathbf{x}_j,\mathbf{x}_k)\}$ is the set of angles, whose pivots are $e_{ij}\in G$. Each angle $(\mathbf{x}_h,\mathbf{x}_i,\mathbf{x}_j,\mathbf{x}_k)$ is modeled by a spring with a resting angle $\overline{a_{ij}}$ and a stiffness coefficient $0<g_{ij}<1$. When $g_{ij}\to 1$, the angle has greater tendency to bend towards $\overline{a_{ij}}$; when $g_{ij}\to 0$, the angle is less likely to do so.

\begin{figure}[H]
 \centering
 \includegraphics[width=0.2\textwidth]{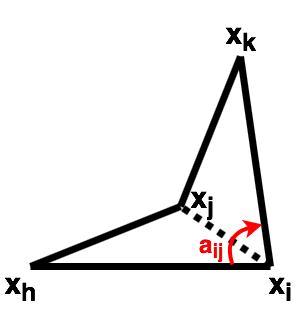}
 \caption{The illustration of angle $(\mathbf{x}_h, \mathbf{x}_i, \mathbf{x}_j , \mathbf{x}_k)$.}
 \label{figure: angle2}
\end{figure}

To translate an acute angle $a_{ij}$ in Figure \ref{figure: angle2} into the coordinates of the vertices $(\mathbf{x}_h,\mathbf{x}_i,\mathbf{x}_j,\mathbf{x}_k)$, we use the following equation:
\begin{equation}
a_{ij}=\arccos\left(\frac{\left(\overrightarrow{\mathbf{x}_h\mathbf{x}_i}\times\overrightarrow{\mathbf{x}_h\mathbf{x}_j}\right)\cdot \left(\overrightarrow{\mathbf{x}_k\mathbf{x}_i}\times\overrightarrow{\mathbf{x}_k\mathbf{x}_j}\right)}{\|\overrightarrow{\mathbf{x}_h\mathbf{x}_i}\times\overrightarrow{\mathbf{x}_h\mathbf{x}_j}\|\cdot\|\overrightarrow{\mathbf{x}_k\mathbf{x}_i}\times\overrightarrow{\mathbf{x}_k\mathbf{x}_j}\|}\right).
\label{eqn:angle}
\end{equation} 

Notice that the above formula only works for $0\le a_{ij}\le\pi$. Therefore, we should preassign a \textbf{Mountain and Valley option} to each angle, where a Mountain crease means $\pi\le a_{ij}\le2\pi$ and a Valley crease means $0\le a_{ij}\le\pi$. For a Mountain crease, if $\overrightarrow{\mathbf{x}_i\mathbf{x}_j}$ is in the same direction as $\overrightarrow{\mathbf{x}_h\mathbf{x}_i}\times\overrightarrow{\mathbf{x}_h\mathbf{x}_j}$, we hold the result $a_{ij}$ calculated by equation \ref{eqn:angle}; otherwise, the actual angle should be $2\pi-a_{ij}$. The conditions are the opposite for a Valley crease. Since the vertex order of each angle matters, we should be careful when we input the angles in practice. For example, one possible angle set for the origami shown in Figure \ref{figure: tetrahedron} could be $\{(\mathbf{x}_1,\mathbf{x}_2,\mathbf{x}_4,\mathbf{x}_5),(\mathbf{x}_3,\mathbf{x}_5,\mathbf{x}_2,\mathbf{x}_4),(\mathbf{x}_6,\mathbf{x}_4,\mathbf{x}_5,\mathbf{x}_2)\}$, where each angle has a M/V option.

\item $\widetilde{G}:=\{(\widetilde{\mathbf{x}}_i,\widetilde{\mathbf{x}}_j)\}$ and $\widetilde{A}:=\{(\widetilde{\mathbf{x}}_h,\widetilde{\mathbf{x}}_i,\widetilde{\mathbf{x}}_j,\widetilde{\mathbf{x}}_k)\}$ are the sets of assistant edges and angles respectively, which are used to model the stiffness of each face. 
$\widetilde{e}_{ij}$ is the selected diagonal of its corresponding face, and $\widetilde{e}_{ij}$ is the pivot of the face bending angle $(\widetilde{\mathbf{x}}_h,\widetilde{\mathbf{x}}_i,\widetilde{\mathbf{x}}_j,\widetilde{\mathbf{x}}_k)$. A stiffness coefficient $0<\widetilde{k}_{ij}<1$ is assigned to each $\widetilde{e}_{ij}$, and $0<\widetilde{g}_{ij}<1$ to each angle $(\widetilde{\mathbf{x}}_h,\widetilde{\mathbf{x}}_i,\widetilde{\mathbf{x}}_j,\widetilde{\mathbf{x}}_k)$. 

Notice that an $n$-vertex face needs $2n-6$ assistant edges and angles, and the combination of assistant edges and angles is not unique. However, the test shows that on a quadrilateral face, the combination of one assistant edge and one corresponding assistant angle yields the most stable simulation, that is the face is less likely to bend itself under the same coefficients. Therefore, this combination is chosen for all experiments with quadrilateral faces in this report. More tests for choosing assistant edges and angles on the faces with vertices more than four is useful, but not necessary in this report.
\end{itemize}

Now, we are able to define an origami structure as a set of $\{G,A,\widetilde{G},\widetilde{A}\}$ together with all the coefficients $\{k,a,\widetilde{k},\widetilde{a}\}$, the resting lengths and angles. Notice that all the angles and edge lengths can be translated into the coordinates of vertices. Therefore, we define the energy of an origami structure as a function of vertices, which is
\begin{equation}
E(\mathbf{x})=E_{edge}(\mathbf{x})+E_{angle}(\mathbf{x})+E_{assistant\,edge}(\mathbf{x})+E_{assistant\,angle}(\mathbf{x}).
\label{eqn:energy}
\end{equation}
As we model each edge and angle by a spring, the energy of each origami structure can be written as a sum of spring energy functions. For example, the energy of edges is 
\[E_{edge}(\mathbf{x})=\frac{1}{2}\sum_{(\mathbf{x}_i,\mathbf{x}_j)\in G}
k_{ij}(\|\mathbf{x}_i-\mathbf{x}_j\|-\overline{l_{ij}})^2.
\]
\end{subsection}

\begin{subsection}{Dynamic Origami Folding Model}\label{subsec:ODE}

In subsection \ref{subsec:defn}, we have defined an origami structure and its energy function. Now, to fold an origami structure in a way such that the energy function decreases, the following ordinary differential equation with respect to time should be solved.
\begin{equation}
\frac{d\mathbf{x}}{dt}= -\frac{\nabla E(\mathbf{x})}{\gamma}+\delta\eta + F_{external},
\label{eqn:ode}
\end{equation}
where $\gamma$ is the parameter of friction. Here, we add a stochastic term $\eta$, which is the white noise $dW_t/dt$, to model the random forcing it feels if it is in a fluid, or subject to random forcing in some other way (random turbulent heat flows, jiggling or shaking, etc). The coefficient $\delta$ is defined as $\sqrt[]{2\beta^{-1}\gamma^{-1}}$. $\beta$, the parameter of inverse temperature, is $1/(kB\cdot T)$, where $T$ is temperature and $kB$ is Boltzmann’s constant. Besides the physical meaning of the stochastic term, $\eta$ also helps the numerical method to traverse more possible energy states before getting stuck in some local minimum.

While solving the ODE system shown in equation \ref{eqn:ode}, we discretize the time space into $N_t=T/\Delta t$ steps and update the vertices in each time step by
\begin{equation}
\mathbf{x}_{j+1}=\mathbf{x}_j-\nabla E(\mathbf{x}_j)\Delta t+\xi+F_{external}\Delta t,
\label{eqn:step}
\end{equation}
where $\xi=\int_{t_j}^{t_{j+1}}\delta dW_t=N(0,\delta^2\Delta t)$. In each simulation, we define a fault-tolerance parameter $\epsilon$, and when $\nabla E(\mathbf{x})<\epsilon$, we believe that the origami structure has reached its final state. Calculating the energy gradients of the length functions is fairly straightforward; however, the gradients of the angle energy is quite involved. More details about calculations are illustrated in appendix \ref{apx:angle} at the end of the report. 

\end{subsection}

\begin{subsection}{Robustness Test and Modifications of the Model}\label{subsec:robust}

So far, we have determined the model of folding a self-folding origami. Before analyzing the results, we show that the discrete numerical method, illustrated by equation \ref{eqn:step}, of solving ODE system in equation \ref{eqn:ode} is robust. Notice that the folding method should be an \textbf{ergodic} dynamical system, which means that the mechanism has the same behavior averaged over time as averaged over all of the space states. Therefore, we compare the energy ensemble \textbf{$E_{ensemble}$} at time $T$, which is the ensemble average over many realizations, and \textbf{$\frac{1}{T}\int_0^TE(t)dt$}, which is the average over long time for one realization. 
\begin{figure}[H]
 \centering
 \includegraphics[width=0.7\textwidth]{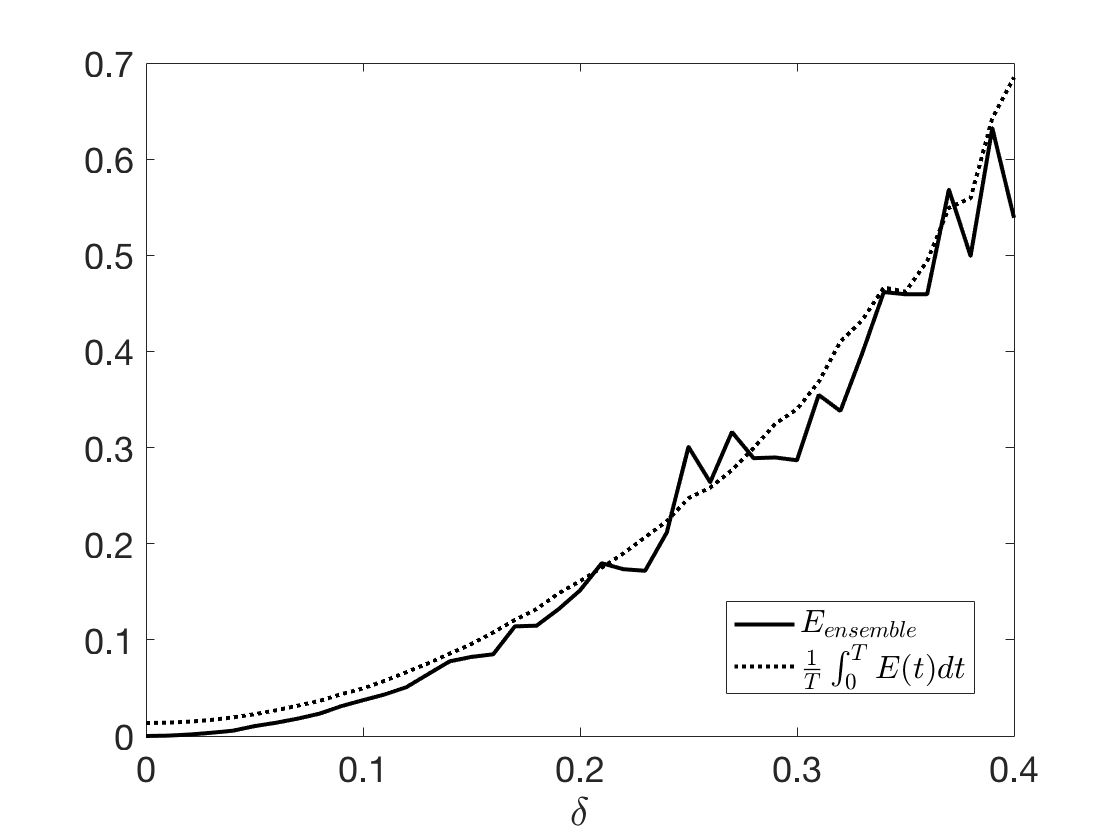}
 \caption{$E_{ensemble}$ and $\frac{1}{T}\int_0^TE(t)dt$ of folding a tetrahedron vs $\delta$ graphs for $50$ realizations, $T=100$, $dt = 0.2$. The unfolded origami is shown in Figure \ref{figure: tetrahedron}, with constant resting edge length $1$ and constant target angle $\arccos\frac{1}{3}$}
 \label{figure: ergodic}
\end{figure}
Figure \ref{figure: ergodic} compares the ensemble energy with $\frac{1}{T}\int_0^TE(t)dt$ at time $T=100$, when folding a tetrahedron under different $\delta$ conditions, that is applying different coefficients on the stochastic term. The plot shows that the ensemble energy and $\frac{1}{T}\int_0^TE(t)dt$ are close to each other in an allowable error scope, which gives us confidence on the ergodicity of the algorithm and it also makes sense that the ensemble energy is increasing with $\delta$. Further tests might be useful, but might not be necessary. 

During the experiments of folding an origami structure, one of the crucial challenges is that the folding mechanism cannot prevent face crossing. We have tried two coping approaches: adding a backtracking line-search method to the algorithm, and modifying the angle function to be continuous. 

\begin{subsubsection}{Backtracking Line-search Method}

Based on the definition of angles, we notice that the face crossing leads to an energy jump, because the angle function is not continuous at zero degree: $0^-$ is $2\pi$ in the angle function. Taking the advantage of this property, the backtracking line-search modifies the step length of the steepest gradient descent, and tells the algorithm to stop folding when two adjacent faces touch each other. At each time point $t_k$, the method will
\begin{itemize}
\item Start with large step length $\alpha_k^0=1$.
\item If $E(\mathbf{x}_k+\alpha_k\Delta\mathbf{x}_k\Delta t)<E(\mathbf{x}_k)$, the algorithm accepts the step length $\alpha_k$.
\item Otherwise, compute $\alpha_k^{i+1}=\rho\alpha_k^i$ with $\rho=\frac{1}{2}$ and go back to the previous step. The algorithm stops folding when $\alpha_k^{i+1}<\epsilon_\alpha$.
\end{itemize}
Therefore, the backtracking line-search method helps us to reach a state on the pathway towards a local minimum of the energy function before stepping beyond the constraint of no face crossing. 
\end{subsubsection}

\begin{subsubsection}{Continuous Angle Function}

Besides the energy local minimums with no face crossing, we are also interested in the energy states without constraints, although the folding process might not be realizable in practice. This goal motivates us to make the angle function continuous.

Firstly, we calculate $\widehat{a}^k_{ij}$ according to the vertices coordinates at the current time point $t_k$. Then we choose $a^k_{ij}$ such that $$a^k_{ij}=\argminA_{a^k_{ij}\in\{\widehat{a}^k_{ij}+2n\pi|n\in\mathbb{Z}\}}\|a^k_{ij}-a^{k-1}_{ij}\|.$$ 
Note that the resting angle $\overline{a_{ij}}$ should be modified in the same way, that is
$$\overline{a^k_{ij}}=\argminA_{\overline{a^k_{ij}}\in\{\overline{a^{k-1}_{ij}}+2n\pi|n\in\mathbb{Z}\}}\|a^k_{ij}-\overline{a^k_{ij}}\|.$$ 
Additionally, the Mountain and Valley option of each angle $a_{ij}$ is checked at each time point, because the modification of the angle function could change the sign of $\nabla E_{a_{ij}}(\mathbf{x})$. Since the sign of $\nabla E_{a_{ij}}(\mathbf{x})$ decides whether two faces linked by $e_{ij}$ tend to attack or repel each other, we modify the energy gradient such that whenever two adjacent faces go across each other, the energy gradient of the angle changes its sign, and therefore the two faces tend to go back in the next time step.
Based on the above modifications on the model, the algorithm can search for local minimums of the energy function with no constraints. All the results presented in section \ref{sec:results} are provided by the algorithm with a continuous angle function.
\end{subsubsection}

\end{subsection}

\begin{subsection}{Introduction to Miura-Ori Crease Pattern} \label{subsec:MO}

	\begin{figure}[H]
	 \centering
	 \includegraphics[width=0.4\textwidth]{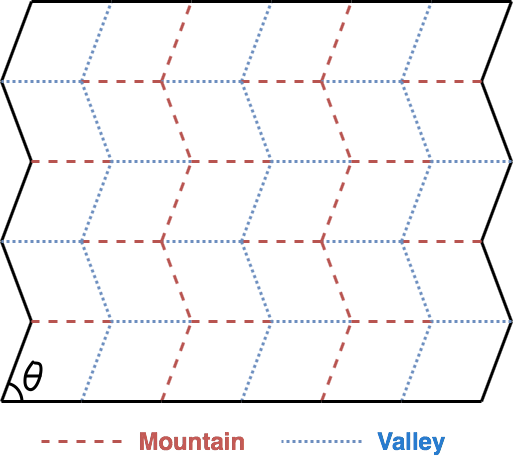}
	 \caption{The crease pattern and Mountain and Valley assignment of a 5 by 6 Miura-Ori and the Miura-Ori angle $\theta$.}
	 \label{figure: Miura-Ori}
	\end{figure}

	In this study, the analysis and tests are based on the case of \textbf{Miura-Ori}, which is a method of folding a flat surface into a smaller area. The crease patterns of the Miura-Ori origami form a tessellation of the surface by parallelograms, which is a collinear quadrilateral metasheet shown in Figure \ref{figure: Miura-Ori}. In a Miura-Ori crease pattern, each of the zigzag paths of creases consists solely of mountain folds or of valley folds, with mountains alternating with valleys from one zigzag path to the next. Each of the straight paths of creases alternates between mountain and valley folds. \cite{wiki} And the Miura-Ori angle $\theta$ is the interior angle of each parallelogram, which is specified in Figure \ref{figure: Miura-Ori}. Notice that it is the M/V assignment that makes a Miura-Ori, so a simply way to check if an origami structure folds in a Miura-Ori way is to check if the M/V assignments of its creases are consistent to the assignment of Miura-Ori.
	 
\end{subsection}

\section{Results} \label{sec:results}
In section \ref{sec:model}, we have introduced the definition of an origami structure and the model of simulating the folding process. Remember that we have introduced the folding mechanisms in section \ref{sec:intro}. There are different methods to install a dynamic ``motor'' on a crease, such that the crease is able to self-fold into the target state. The imposed strain on a crease implies that the angle coefficient of this crease is positive, and a crease without imposed force means that its angle coefficient is zero. The widely-used model of folding a self-folding origami imposes strain on all the creases in its crease pattern, that is the imposed pattern of strain assignments is exactly the crease pattern of this origami structure. But in fact, we could fold a collinear quadrilateral metasheet into Miura-Ori folded state with only part of its creases pinched. Then it is natural to define an \textbf{optimal pattern of strain (OPS)}, which records the minimum ``functional'' creases that enable the metasheet to fold into the desired Miura-Ori configuration. In this section, we address the following three questions introduced in section \ref{sec:intro}: Does OPS exist? Why it works? And how to determine the OPS using only local information of the initial flat state? 

\begin{subsection}{Question 1: Does OPS Exist?}\label{sec:results1}
	
In this subsection, we show that there exists an optimal pattern of strain that can fold a collinear quadrilateral metasheet into a Miura-Ori configuration stably. The stability of a particular pattern of creases is determined by the probability of success, i.e. the probability of successfully folding a flat metasheet into a Miura-Ori folded state with only these creases pinched. Notice that there are $m$ ($m=$ number of creases) one-crease-patterns, $(_{\,2}^m)=\frac{m\times (m-1)}{2!}$ two-crease-patterns, etc. Start from each one-crease-pattern, we apply the following test and calculate the probability of success.
\begin{itemize}
	\item Test Design\\
	In each realization, the collinear quadrilateral metasheet starts from a flat state with constant resting angles $\arccos (0.95)$, and constant angle coefficients $0.05$ for those creases in the pattern of strain (angle coefficient is zero if the crease is not pinched). When folding the origami, we add large noise to the system and allow the origami to fold for sufficiently long time $T$ ($dt = 0.1$). We repeat this test for hundreds of realizations and calculate the probability of success using following metrics.
	\item Metric of Success
	\begin{itemize}
		\item Metric I: the folded state is a Miura-Ori if the M/V assignments of its creases are consistent to the assignment of Miura-Ori.
		\item Metric II: the folded state is a Miura-Ori if $\sum_{i\in\text{crease set}}\|\theta_i-\bar\theta_i\|^2<\text{threshold}$, where $\theta_i$ is the current angle of crease $i$ and $\bar\theta_i$ is the reseting angle of crease $i$. The threshold depends on the size of the metasheet.
		\item Metric III: the folded state is a Miura-Ori if it satisfies both Metric I and II. This is the default metric we use.
	\end{itemize}
\end{itemize}

If more than about $95\%$ of the realizations end up with Miura-Ori, then this pattern of creases is a possible OPS. Among all possible OPS with same number of creases, we choose the one with highest stability. The OPS may not be unique, if more than one patterns with same number of creases can do the job with equal stability. If no one-crease-pattern can fold the origami to Miura-Ori stably, we continue to test two-crease-patterns and so on. 

\begin{figure}[H]
    \centering
    \begin{subfigure}[t]{0.5\textwidth}
        \centering
        \includegraphics[width=0.5\textwidth]{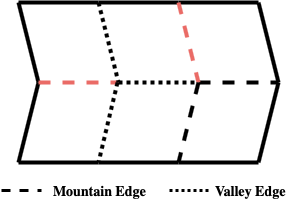}
    \end{subfigure}%
    ~ 
    \begin{subfigure}[t]{0.5\textwidth}
        \centering
        \includegraphics[width=0.5\textwidth]{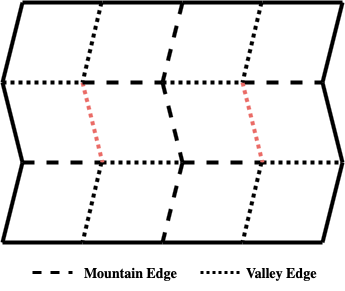}
    \end{subfigure}
	\caption{The crease pattern and Mountain and Valley assignments of a 2 by 3 (left) and a 3 by 4 (right) Miura-Ori. Red creases represent the OPS we find.}
	\label{figure: 2by33by4}
\end{figure}	

For example, Figure \ref{figure: 2by33by4} shows the OPS we find through the above numerical experiments of a 2 by 3 metasheet and a 3 by 4 metasheet. The Miura-Ori angle for both case is $80$ degrees. Pinching the creases in the OPS of a 2 by 3 metasheet and adding a large noice with noise coefficient $\delta = 0.1$, 200 out of 200 realizations end up with Miura-Ori folded state using any of the three metrics, i.e. the probability of success is $100\%$. In the case of the 3 by 4 metasheet, the probability of success is $92.4\%$ (924 out of 1000) if $\delta = 0.1$ and $99.9\%$ (999 out of 1000) if $\delta = 0.0753$. In fact, $95\%$ is not a strict cut-off line to determine OPS is stable or not, because this probability depends on not only the determination of OPS, but also other factors such as the noise and Miura-ori angle. Figure \ref{figure: contour} shows the probability of ending up with Miura-ori configuration pinching the OPS we find on 3 by 4 metasheet. We can see that the OPS is more stable when we add smaller noise to the system and our OPS works better when Miura-ori angle is between $70-80$ degrees. Further tests are recommended but not necessary in this case, because we have already showed that there exists an optimal pattern of strain, with which we can stably fold a metasheet into Miura-Ori folded state pinching least creases.

\begin{figure}[H]
    \centering
	\begin{adjustwidth}{-25mm}{}
    \includegraphics[width=1.35\textwidth]{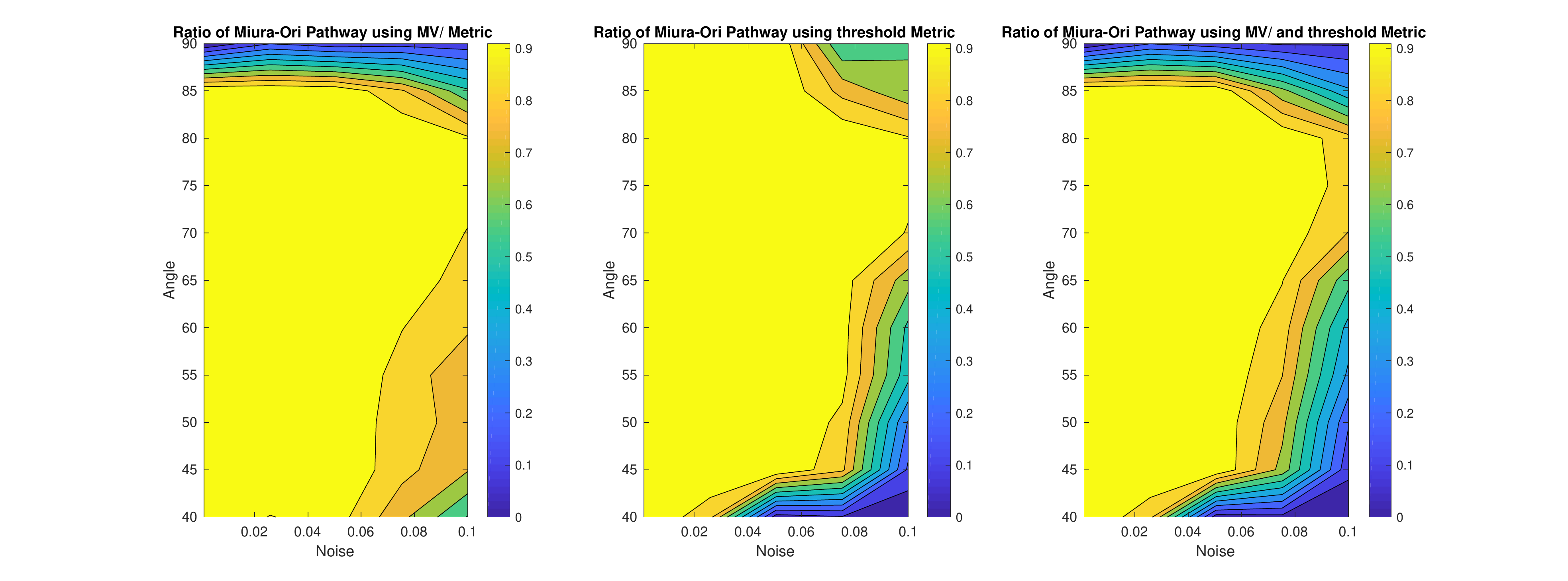}
	\end{adjustwidth}
	\caption{The probability of ending up with Miura-ori configuration pinching the OPS of 3 by 4 metasheet with respect to Metric I (left), Metric II (middle), and Metric III (right).}
	\label{figure: contour}
\end{figure}	
	
\end{subsection}

\begin{subsection}{Question 2: Why the OPS Can Truely Fold a Flat Metasheet into Miura-Ori State?}\label{sec:results2}
In section \ref{sec:results1}, we show the existence of the optimal pattern of strain. But why the OPS works? Why can we fold a metasheet into Miura-Ori stably by pinching only the OPS? In this section, we explain it by comparing the energy evolution along each possible folding pathway. So firstly, in section \ref{subsec:DFS}, we introduce how we find the possible folded states using a Depth First Search algorithm. And in section \ref{subsec:energy}, we figure out that the energy predominance of the desired Miura-Ori pathway during the initial period of time accounts for why OPS works. 
	
\begin{subsubsection}{Possible Folded States Determined by a Set of Crease-Pinchings}\label{subsec:DFS}
	\begin{figure}[H]
	 \centering
	 \includegraphics[width=0.15\textwidth]{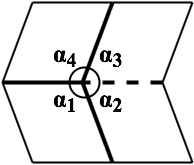}
	 \caption{A degree-four vertex.}
	 \label{figure: vertex}
	\end{figure}
In this subsection, we introduce our method of finding the available folded states determined by a set of crease-pinchings. In fact, this is the same as finding the available M/V assignments, if we have preset the M/V assignments on the creases we pinch. First, we look at the constraints of the M/V assignments on an essential component of a metasheet, a \textbf{degree-four vertex}. As shown in Figure \ref{figure: vertex}, a degree-four vertex is four rigid faces attached by four edges that meet a vertex. We specify the flat-state geometry by the set of sector angles $\{\alpha_i\}$, where $0<\alpha_i<\pi$ and $\sum_i\alpha_i=2\pi$. Huffman \cite{Huffman} noted that one folding angle must have the opposite Mountain and Valley option from the rest. Additionally, Waitukaitis \cite{Scott} showed that edge $e_{ij}$ can be the ``unique'' fold with the M/V option opposite from the rest, only if the sum of sector angles of the two plates connected by edge $e_{ij}$ is less than or equal to $\pi$. For example, the case shown in Figure \ref{figure: vertex} is an available M/V assignment, since $\alpha_2+\alpha_3\le\pi$.
	\begin{figure}[H]
	 \centering
	 \includegraphics[width=0.4\textwidth]{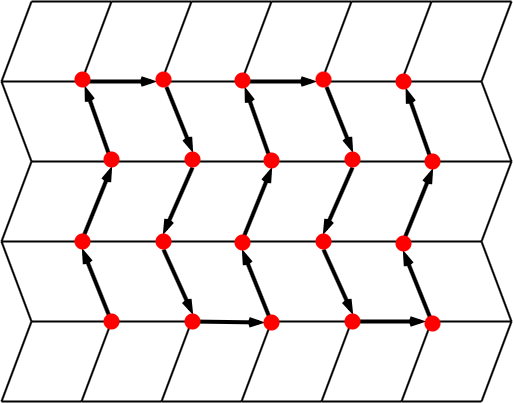}
	 \caption{A directed graph of degree-four vertices on a 5 by 6 metasheet.}
	 \label{figure: DFS}
	\end{figure}

	Based on the above two constraints on the M/V assignments at a degree-four vertex, we design a \textbf{Depth First Search (DFS)} algorithm to traverse a directed graph, which is made of degree-four vertices on a metasheet (Figure \ref{figure: DFS}). At each vertex, the algorithm works out the available M/V assignments that satisfy the constraints, and then calls next vertex if there are any.
	\begin{table}[H]
	\centering
	\begin{tabular}{c|*6c}
	\toprule
	\textbf{} & \textbf{2} & \textbf{3} & \textbf{4} & \textbf{5} & \textbf{6}\\
	\hline
	\textbf{2}&6&18&54&162&486\\
	\textbf{3}&18&82&374&1706&7782\\
	\textbf{4}&54&374&2604&18150&126534\\
	\textbf{5}&162&1706&18150&193662&2068146\\
	\textbf{6}&486&7782&126534&2068146&33865632\\
	\bottomrule
	\end{tabular}
	\caption{The number of available M/V assignments of a collinear quadrilateral metasheet of different sizes. The entry in row $i$ and column $j$ is the number of available M/V assignments of a $i$ by $j$ collinear quadrilateral metasheet.}
	\label{table:DFS}
	\end{table}
	Table \ref{table:DFS} shows the number of available M/V assignments on a collinear quadrilateral metasheet of difference sizes. It makes sense that the number of available assignments grows exponentially with the size of metasheets. 
\end{subsubsection}

\begin{subsubsection}{Energy Evolution on Possible Folding Pathways}\label{subsec:energy}
For simplicity, we take a small 2 by 3 collinear metasheet as an example and the OPS we refer in this subsection is the red creases shown in Figure \ref{figure: 2by33by4}. Using the method introduced in section \ref{subsec:DFS}, we find that there are four possible folded states of a 2 by 3 metasheet with M/V assignments consistent to the OPS shown in Figure \ref{figure: states}. The state 4 is the Miura-Ori configuration, which is our desired folded state.

\begin{figure}[H]
 \centering
 \includegraphics[width=0.7\textwidth]{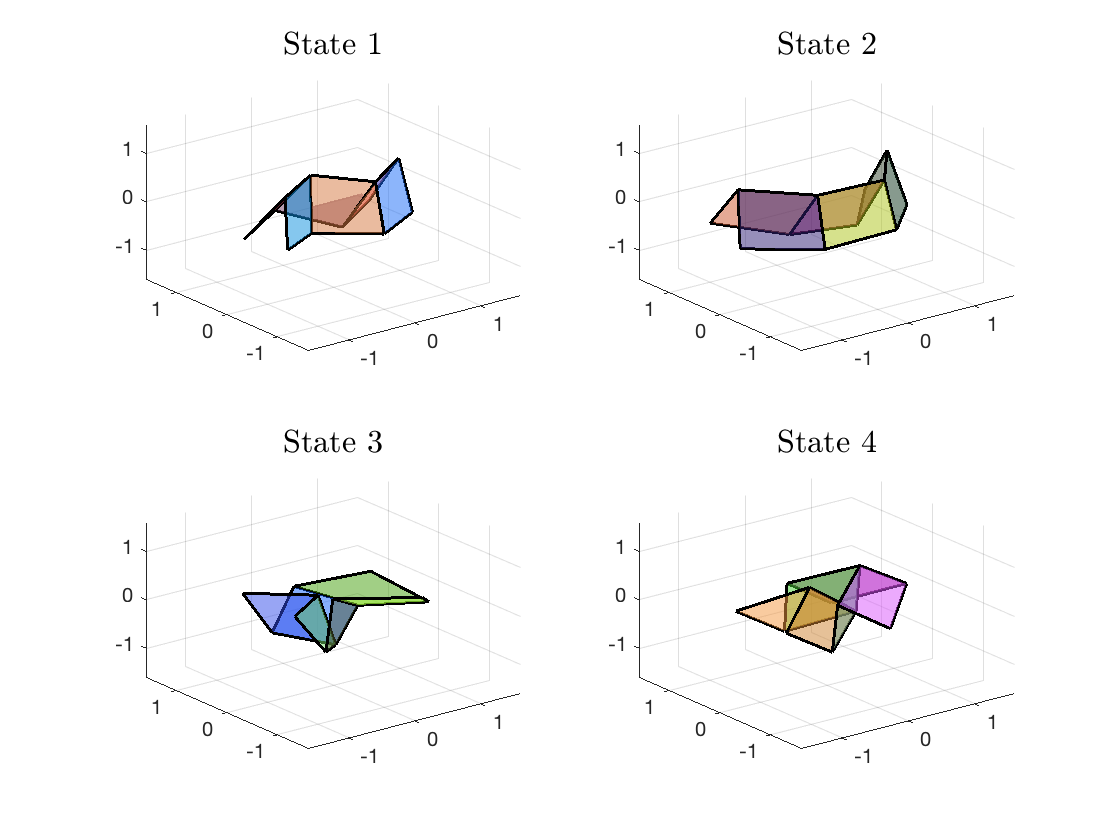}
 \caption{Four available folded configurations of a 2 by 3 collinear quadrilateral metasheet with Mountain and Valley assignment consistent to the optimal pattern of strain.}
 \label{figure: states}
\end{figure}

To compare the energy evolution along these four pathways, we firstly define two energy functions.
\begin{itemize}
	\item $E_{optimal}$ for $i=1\cdots 4$: the energy function which looks at only the angles of creases in OPS (and all the edges). That is, the angles of the creases that are not in OPS are ignored and have zero coefficients.
	\item $E^{(i)}_{state}$ for $i=1\cdots 4$: the energy function that takes account of all the angles (and all the edges).
\end{itemize}

Figure \ref{figure: energy} and \ref{figure: gradient} plot the average $E_{optimal}$ and the average gradient of $E_{optimal}$, which is $$\frac{\nabla E_{desired}\cdot\nabla E_{temporary}^{(i)}}{\|\nabla E_{temporary}^{(i)}\|},$$ over 100 realizations on the pathway of each folded state $i$. When $t$ is less than $15$ or so, $E_{optimal}$ on the pathway of Miura-ori, which is state $4$, is the lowest among the four pathways. It makes sense that the energy predominance of the initial period of time may account for the feasibility of the optimal pattern of strain, since after $t=15$, the metasheet has already been stuck in its Miura-Ori pathway.

\begin{figure}[H]
 \centering
 \includegraphics[width=0.7\textwidth]{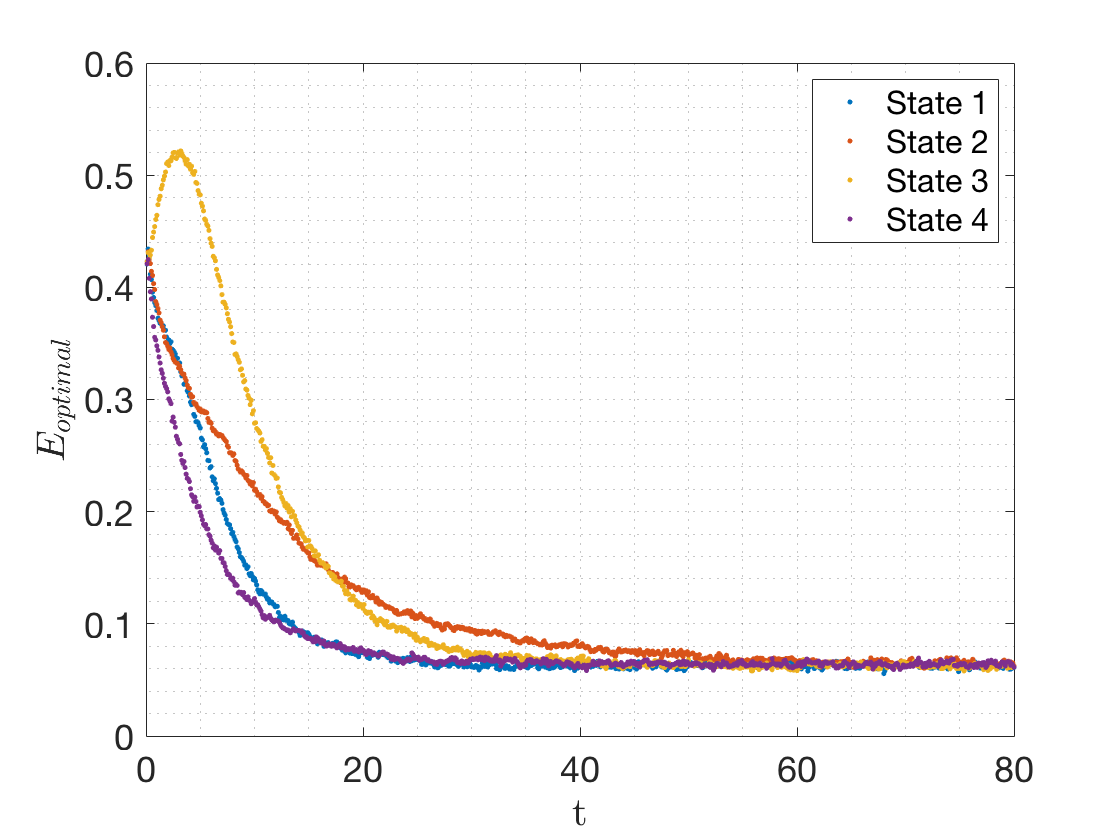}
 \caption{$E_{optimal}$ along pathway $i$ for $i=1,\cdots,4$ vs time $t$ graphs for constant resting angles $\arccos{(0.95)}$, and constant angle coefficients $0.05$ with $0 < t < 80$, $dt = 0.1$, $\delta = 0.1$.}
 \label{figure: energy}
\end{figure}

Furthermore, the energy evolution of the pathway to state $3$ is remarkable. The energy jump in the initial period of time represents the order of the folding process. That is we need to first fold horizontal creases and then vertical creases to end up with state $3$. The low average gradient of $E_{optimal}$ demonstrates that the origami has to fold in a pathway violating the direction of $E_{optimal}$ during the initial period of time. The above evidence explains why state $3$ is the least favorable folded state. 
\begin{figure}[H]
 \centering
 \includegraphics[width=0.7\textwidth]{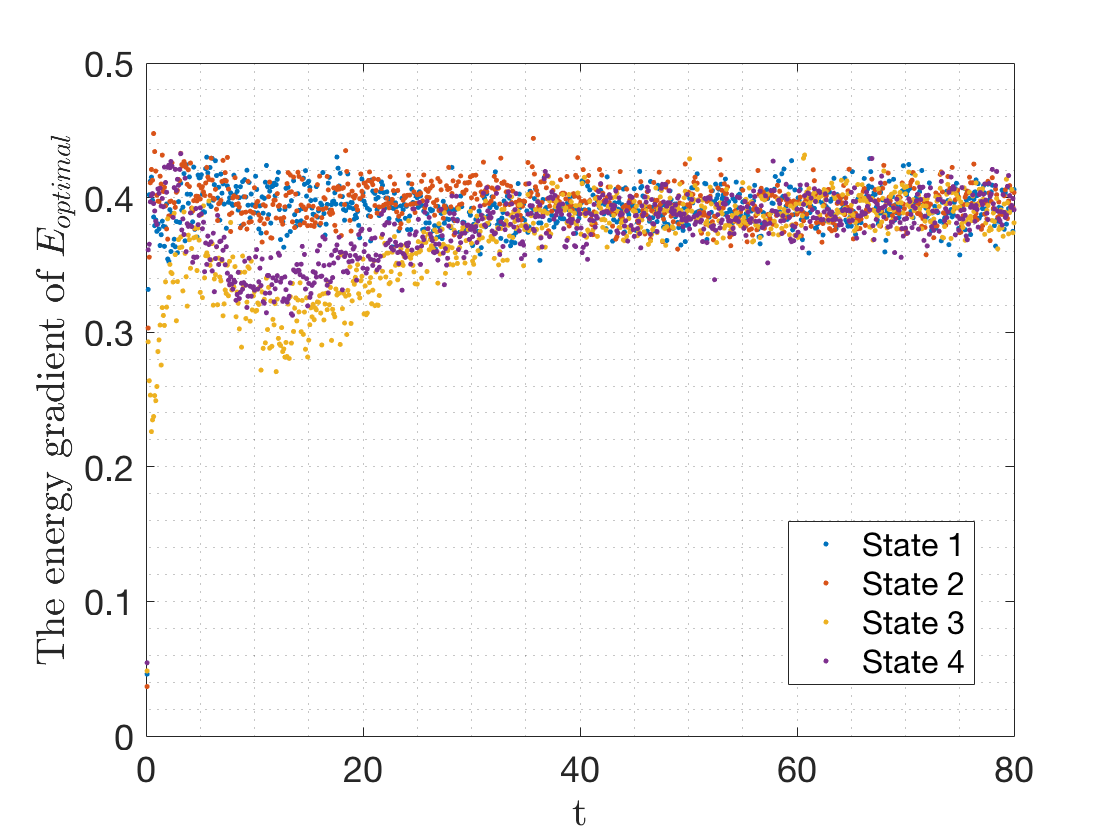}
 \caption{The energy gradient $\frac{\nabla E_{desired}\cdot\nabla E_{temporary}^{(i)}}{\|\nabla E_{temporary}^{(i)}\|}$ along pathway $i$ for $i=1,\cdots,4$ vs time $t$ graphs for constant resting angles $\arccos{(0.95)}$, and constant angle coefficients $0.05$ with $0 < t < 80$, $dt = 0.1$, $\delta = 0.1$.}
 \label{figure: gradient}
\end{figure}

\end{subsubsection}

\end{subsection}

\begin{subsection}{Question 3: Can We Determine the OPS using Only Local Information of the Initial Flat State?}\label{sec:results3}
	
	In section \ref{sec:results1}, we introduce how we find the OPS using the probability of success after a sufficiently long period of time. However, in section \ref{sec:results2}, we find that the origami actually gets stuck in its folding pathway after only a short period of time. So is it possible to predict the OPS, or at least some feastures of the OPS, using only local information of the initial flat state? Notice that the OPS of a 3 by 4 metasheet only contains two vertical creases, but the unique structure of the collinear quadrilateral metasheet propogates the force of these two creases to the entire metasheet and fits it in a Miura-Ori track. So in this section, we address the following questions: \textbf{Does the force generated by the OPS at the initial state actually push the origami to a Miura-Ori pathway? And are there any other pattern of strains that could do the same job? If so, are these patterns of strains also OPS, and does they provide any useful information on the determination of the OPS?} In section \ref{subsec:rigidity}, we introduce how we calculate the projected force on the metasheet caused by certain crease pinchings. And in section \ref{subsec:insights}, we discuss the above questions.
	
\begin{subsubsection}{The Rigidity Matrix and the Projected Force}\label{subsec:rigidity}
Let's firstly clarify some notations.
\begin{itemize} 
	\item $X(t)=(x_1^x,x_1^y,x_1^z,x_2^x,\cdots,x_n^z)\in\mathbb{R}^{3n}$ denotes the coordinates of all the vertices as a functino of time. Similarly, $X_i(t)=(x_i^x,x_i^y,x_i^z)$ for $i=1,\cdots, n$, is the coordinate of one vertex. 
	\item $V=(v_1^x(0),v_1^y(0),v_1^z(0),v_2^x(0),\cdots,v_n^z(0))\in\mathbb{R}^{3n}$ denotes the velocity of the coordinates of all the vertices at the initial state $t=0$. That is, $\left.\frac{d}{dt}\right|_{t=0}X(t)=V$. And $V_i=(v_1^x(0),v_1^y(0),v_1^z(0))$ for $i=1,\cdots, n$, is the initial velocity of one vertex. 
\end{itemize}
	Notice that the constraints of $X$ at the initial state are the fixed length and angles of the edges, assistant edges, and assistant angles. More specifically, the following equations should be satisfied.
		\begin{equation*}
		\begin{cases}
			|X_i(t)-X_j(t)|^2=e_{ij}^2, &\text{for }(X_i,X_j)\in G\text{ and }\widetilde G,\\
			a_{i,j}(X(t),t)=\overline a_{i,j}, &\text{for }(X_i,X_j)\in \widetilde A.
		\end{cases}
		\end{equation*}
Taking the derivative of both sides at $t=0$, we get the constraint of $V$:
			\begin{equation}
			\begin{cases}
				[X_i(0)-X_j(0)][V_i-V_j]=0, &\text{for }(X_i,X_j)\in G\text{ and }\widetilde G,\\
				(\nabla_X a_{i,j}) V=0, &\text{for }(X_i,X_j)\in \widetilde A.
			\end{cases}
			\label{eqn:constraints}
			\end{equation}
The calculation of the angle gradient at flat state, i.e., $\nabla_X a_{i,j}$ when $a_{i,j}=\pi$, is a bit tricky, because the angle gradient at flat state is in the form of $\frac 00$. Instead, we use an asympotic analysis to calculate the gradient. More details of this calculation are shown in Appendix \ref{apx:asymptotic}. The system of linear equations (\ref{eqn:constraints}) can be written as $R(X(0))V=0$, where $R(X)\in\mathbb{R}^{m\times 3n}$ ($m$ constraints in total) is our rigidity matrix. Each row of the rigidity matrix specified one constraint of $V$ and therefore the null space of $R$ specified the allowed directions that the origami could go to at the initial state.

Let $F:=\sum_{(i,j)\in OPS}\nabla_X a_{i,j}$ denote the force of pinching the creases in OPS, and $T$ denote the orthonormal basis of $Null(R)$. Then the projected force $F_{proj}$ on the metasheet is $F_{proj}=(TT^T)F$.
	\end{subsubsection}
	
\begin{subsubsection}{Discussions on What We Learn from the Projected Force}\label{subsec:insights}
To verify that the OPS we choose indeed fold the origami in Miura-Ori way at the initial state, we use the Metric I specified in section \ref{sec:results1} to check if $X(0)+\epsilon F_{proj}(OPS)$ ($0<\epsilon \leq 1$) is a Miura-Ori state. As we expected, the answer is yes, as long as the Miura-Ori angle is less than 90 degrees. This fact proves why the OPS works in another perspective. However, when the Miura-Ori angle is 90 degrees, the metasheet becomes a tesselation of squares and therefore cannot transfer the force of crease pinchings any more. This fact also explains why the probability of ending up with Miura-Ori shown in Figure \ref{figure: contour} is very low when Miura-Ori angle is 90 degrees. 

	\begin{figure}[H]
	 \centering
	 \includegraphics[width=0.4\textwidth]{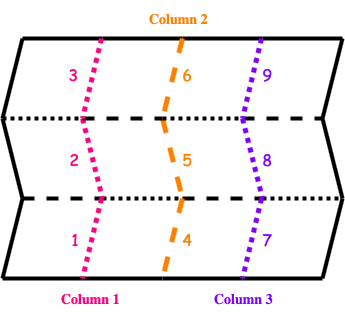}
	 \caption{The crease pattern and Mountain and Valley assignment of a 3 by 4 Miura-Ori.}
	 \label{figure: colmn}
	\end{figure}

It is not surprising that there are other pattern of strains that could push the initial metasheet to a Miura-Ori direction. Our tests show that pinching only one vertical crease from column 2 in Figure \ref{figure: colmn} already generates the force of folding the origami to Miura-Ori pathway. Besides, any two distinct vertical creases from any column can do the same job. Using the tests specified in section \ref{sec:results1}, we calculated the probability of success of these pattern of strains. The tests show that if we pinch creases from the same column (including pinching one crease from column 2) or from adjacent columns (column 1+2 or column 2+3), the probability of success is very sensitive to noise. If the noise is small ($\delta=0.01$), the probability of success is high (mostly more than 80 percents) under Metric I, but not under Metric II. That is, these pattern of strains can truely fold the origami to Miura-Ori, but as time goes, the origami folds more and more slowly, and may take infinite time to finish up with a satisfied Miura-Ori under Metric II. If we pinch one crease from column 1 and one from column 3, the performance gets better than other pattern of strains, but OPS is still the one with highest probability of success. In conclusion, although we cannot determine the OPS directly from the local information of the initial state, we find that vertical creases should be crucial components of OPS, because they generate forces that push the entire metasheet into Miura-Ori track, which horizontal creases cannot. Furthermore, if we have a limited number of crease pinchings, to achieve better stability, the choice of creases should be better separated, not concentrated.
	
\end{subsubsection}

\end{subsection}

\begin{section}{Conclusion}\label{sec:conclu}
In this study, we introduced the model of folding an origami structure. The obstacles of dealing with face crossing is identified and alleviated. We argued that there is an optimal pattern of strain, which promises a collinear quadrilateral metasheet to fold into a Miura-Ori with fewest creases pinched. We showed that the authenticity of this optimal pattern of strain can be confirmed by comparing the energy evolution along the pathway of each possible folded state. Furthermore, we show that vertical creases are crucial for the determination of OPS, because a very few of them can generate forces that push the entire metasheet folding in a Miura-Ori pathway.

Our work suggests many interesting possible directions in expansion of this project. Firstly, although we have verified that a self-folding metasheet could fold into a Miura-Ori under the OPS, the answer is still unknown to the question that why the energy function favors Miura-Ori more than the other states. It is interesting that even though there are several possible local minima the system could go to, it overwhelmingly evolves to just one of these. Why is that? What property of the energy landscape makes this so? Furthermore, what particular structures of the collinear quadrilateral metasheet propogate the force generated by a few crease pinchings? Is there any other structures could we generalize our results? These are all interesting questions to answer in expansion of this project. 

\end{section}

\renewcommand{\abstractname}{Acknowledgements}
\begin{abstract}
I thank Miranda Holmes-Cerfon, Pejman Sanaei, and Jason Kaye for insightful advice and discussions. I acknowledge National Science Foundation (DMS-1646339) for funding and the Courant Institute for computing resources.
\end{abstract}

\begin{appendices}
\begin{section}{Calculation of the Gradient of Angle Energy}\label{apx:angle}

\begin{figure}[H]
 \centering
 \includegraphics[width=0.3\textwidth]{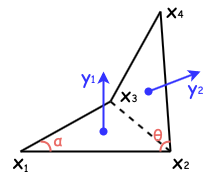}
 \caption{The illustration of angle $(\mathbf{x}_1, \mathbf{x}_2, \mathbf{x}_3 , \mathbf{x}_4)$}
 \label{figure: angle}
\end{figure}
As shown in Figure \ref{figure: angle}, let $\mathbf{y}_1(\mathbf{x})$ be a normal vector of triangle $(\mathbf{x}_1, \mathbf{x}_2 \mathbf{x}_3)$, and $\mathbf{y}_2(\mathbf{x})$ be a normal vector of triangle $(\mathbf{x}_4, \mathbf{x}_2, \mathbf{x}_3)$, written as 
\[ \mathbf{y}_1(\mathbf{x})= \overrightarrow{\mathbf{x}_1\mathbf{x}_2}\times\overrightarrow{\mathbf{x}_1\mathbf{x}_3} \text{, and } \mathbf{y}_2(\mathbf{x})=\overrightarrow{\mathbf{x}_4\mathbf{x}_2}\times\overrightarrow{\mathbf{x}_4\mathbf{x}_3}.
\]
Then the actual angle $a_{x_2,x_3}$ is
\[
a_{x_2,x_3}(\mathbf{x})=\arccos\frac{\mathbf{y}_1(\mathbf{x})\cdot\mathbf{y}_2(\mathbf{x})}{\|\mathbf{y}_1(\mathbf{x})\|\cdot\|\mathbf{y}_2(\mathbf{x})\|}.
\]
The energy of angle $(x_1, x_2, x_3 , x_4)$ is 
\[
E_{a_{x_2,x_3}}(\mathbf{x})=\frac{1}{2}g_{x_2,x_3}\left(a_{x_2,x_3}- \overline{a_{x_2,x_3}}\right)^2,
\]
where $g_{x_2,x_3}$ is the angle coefficient and $\overline{a_{x_2,x_3}}$ is the resting angle. Then we take the gradient of the energy with respect to $x$, and get
\[
\nabla E_{a_{x_2,x_3}}(\mathbf{x})=
-\frac{a_{x_2,x_3}- \overline{a_{x_2,x_3}}}{\sqrt[]{1-\left(\frac{\mathbf{y}_1\cdot\mathbf{y}_2}{\|\mathbf{y}_1\|\|\mathbf{y}_2\|}\right)^2}}
\left[\frac{\mathbf{y}_2\nabla_{\overrightarrow{\mathbf{x}}}\mathbf{y}_1+ \mathbf{y}_1\nabla_{\overrightarrow{\mathbf{x}}}\mathbf{y}_2}{\|\mathbf{y}_1\|\|\mathbf{y}_2\|}
-\mathbf{y}_1\mathbf{y}_2\cdot \left(
\frac{\mathbf{y}_1\nabla_{\overrightarrow{\mathbf{x}}}\mathbf{y}_1}{\|\mathbf{y}_1\|^3\|\mathbf{y}_2\|}+\frac{\mathbf{y}_2\nabla_{\overrightarrow{\mathbf{x}}}\mathbf{y}_2}{\|\mathbf{y}_1\|\|\mathbf{y}_2\|^3}
\right)\right].
\]
\end{section}

\begin{section}{The Asymptotic Analysis of the Gradient of angle $\theta$ when $\theta=\pi$}\label{apx:asymptotic}
The notations of the angle is specified in Figure \ref{figure: angle}. Assume $x_2x_3$ is a Valley edge and let $0<h\ll 1$ denotes a small increment of angle such that current angle is $\theta-h=\pi-h$. Since the Miura-Ori angle is $\alpha$ and the default length of edge is one, then the length of the diagonal is $l=\|x_2x_3\|=\sin(\alpha)/\sin(\frac{\pi-\alpha}{2}).$ Now we can write the coordinates of the four vertices:
\begin{align*}
	x_1=&(-\sqrt{1-l^2/4},0,0),\\
	x_2=&(0,-l/2,0),\\
	x_3=&(0,l/2,0),\\
	x_4=&(\sqrt{1-l^2/4}\cos h,0,\sqrt{1-l^2/4}\sin h)\\
	=&(\sqrt{1-l^2/4}-\frac 12\sqrt{1-l^2/4}h^2,0,\sqrt{1-l^2/4}h)+O(h^3).
\end{align*}
Then the normal vectors are 
\begin{align*}
	y_1=&(0,0,l\sqrt{1-l^2/4}),\\
	y_2=&(l\sqrt{1-l^2/4}h,0,-l\sqrt{1-l^2/4}+\frac l2\sqrt{1-l^2/4}h^2)+O(h^3).
\end{align*}
Plug in the gradient of angle with respect to $(x_1^x,x_1^y,x_1^z,x_2^x,x_2^y,x_2^z,x_3^x,x_3^y,x_3^z)$, we get
\begin{align*}
	\nabla_x\theta=&-\frac{1}{\sqrt{1-\frac{(y_1\cdot y_2)^2}{|y_1|^2|y_2|^2}}}\left[\frac{\nabla y_1\cdot y_2+\nabla y_2\cdot y_1}{|y_1||y_2|}-y_1\cdot y_2\left(\frac{\nabla y_1\cdot y_1}{|y_1|^3|y_2|}+\frac{\nabla y_2\cdot y_2}{|y_1||y_2|^3}\right) \right]\\
	=&(0,0,-\frac{2}{\sqrt{4-l^2}},-\frac{4h}{(4+h^4)\sqrt{4-l^2}},0,\frac{8-2h^2+h^4}{(4+h^4)\sqrt{4-l^2}},-\frac{4h}{(4+h^4)\sqrt{4-l^2}},0,\\
	&\frac{8-2h^2+h^4}{(4+h^4)\sqrt{4-l^2}},\frac{8h}{(4+h^4)\sqrt{4-l^2}},0,\frac{4(-2+h^2)}{(4+h^4)\sqrt{4-l^2}})^T+O(h^3)
\end{align*}
Therefore, when $h\to 0$, the gradient of angle is 
\begin{align*}
	\nabla_x\theta=&(0,0,-\frac{2}{\sqrt{4-l^2}},0,0,\frac{2}{\sqrt{4-l^2}},0,0,\frac{2}{\sqrt{4-l^2}},0,0,-\frac{2}{\sqrt{4-l^2}})^T.
\end{align*}
\end{section}
\end{appendices}


\begin{thebibliography}{9}
\bibitem{Pandey} 
Pandey, Shivendra; Ewing, Margaret; Kunas, Andrew; Nguyen, Nghi; Gracias, David H.; Menon, Govin. Algorithmic design of self-folding polyhedra. \textit{National Academy of Sciences}. \textbf{2011}. doi:10.1073/pnas.1110857108.

\bibitem{Miskin} 
Miskin, Marc Z.; Dorsey, Kyle J.; Bircan, Baris; Han, Yimo; Muller, David A.; McEuen, Paul L.; Cohen, Itai. Graphene-based bimorphs for micron-sized, autonomous origami machines. \textit{National Academy of Sciences}. \textbf{2018}. doi:10.1073/pnas.1712889115.

\bibitem{Plucinsky} 
Plucinsky, Paul; Kowalski, Benjamin A.; White, Timothy J.; Bhattacharya, Kaushik. Patterning nonisometric origami in nematic elastomer sheets. \textit{Soft Matter}. \textbf{2018}. doi:10.1039/C8SM00103K.

\bibitem{Levi} 
Levi H. Dudte; Etienne Vouga; Tomohiro Tachi; L. Mahadevan. Programming curvature using origami tessellations. \textit{Nature Materials}. \textbf{2016}. doi:10.1038/nmat4540.

\bibitem{Matthew} 
Matthew B. Pinson; Menachem Stern; Alexandra Carruthers Ferrero; Thomas A. Witten; Elizabeth Chen; Arvind Murugan. Self-folding origami at any energy scale. \textit{Nature Communications}. \textbf{2017}. doi:10.1038/ncomms15477.

\bibitem{Bryan} 
Bryan Gin-ge Chen; Christian D. Santangelo. Branches of triangulated origami near the unfolded state. \textit{Physical Review. X 8, 011034}. \textbf{2018}. doi:10.1103/PhysRevX.8.011034.

\bibitem{Menachem} 
Menachem Stern; Matthew B. Pinson; Arvind Murugan. The Complexity of Folding Self-Folding Origami. \textit{Phys. Rev. X 7, 041070}. \textbf{2017}. doi:10.1103/PhysRevX.7.041070.

\bibitem{Christian} 
Christian D. Santangelo. Extreme Mechanics: Self-Folding Origami. \textit{Annu. Rev. Condens. Matter Phys}. \textbf{2017.8:165-83}. doi:10.1146/annurev-conmatphys-031016-025316.

\bibitem{Scott} 
Scott Waitukaitis; Rémi Menaut; Bryan Gin-ge Chen; Martin van Hecke. Origami Multistability: From Single Vertices to Metasheets. \textit{Phys. Rev. Lett. 114, 055503}. \textbf{2015}. doi:10.1103/PhysRevLett.114.055503.

\bibitem{Huffman} 
David A. Huffman. Curvature and Creases: A Primer on Paper. \textit{IEEE Transactions on Computers, Vol C-25, 1010}. \textbf{1976}.

\bibitem{wiki}
Wikipedia contributors. (2018, March 9). Miura fold. In Wikipedia, The Free Encyclopedia. Retrieved 03:01, August 9, 2018, from \texttt{https://en.wikipedia.org/w/ index.php?title=Miura\_fold\&oldid=8296357174}

\end{thebibliography}
\end{document}